\documentclass[twocolumn,amsmath,amssymb]{revtex4}
\usepackage{dcolumn}
\usepackage{bm}
\usepackage{graphicx}
\usepackage{subfigure}
\usepackage{color}

\usepackage[normalem]{ulem}
\usepackage{color}
\usepackage{hyperref}

\newcommand{\tcb}{\textcolor{blue}}

  \newcommand{\jy}[1]{\tcb{\bf {[[Jiayue: #1]]}}}

\begin{document}
\title{Critical slowing down of black hole phase transition and kinetic crossover in supercritical regime}

\author{Ran Li$^a$}
\thanks{liran@qfnu.edu.cn, equal contribution.}

\author{Kun Zhang$^b$}
\thanks{zhangkun@ciac.ac.cn, equal contribution.}

\author{Jiayue Yang$^{c,d}$}
\thanks{j43yang@uwaterloo.ca}

\author{Robert B. Mann$^{c,d}$}
\thanks{rbmann@uwaterloo.ca}

\author{Jin Wang$^e$}
\thanks{jin.wang.1@stonybrook.edu, Corresponding author.}

\affiliation{$^a$Department of Physics, Qufu Normal University, Qufu, Shandong 273165, China}

\affiliation{$^b$ State Key Laboratory of Electroanalytical Chemistry, Changchun Institute of Applied Chemistry, Chinese Academy of Sciences, Changchun, China, 130022}

 \affiliation{$^c$Department of Applied Mathematics and Department of Physics and Astronomy, University of Waterloo,
200 University Avenue West, Waterloo, ON, N2L 3G1, Canada}

\affiliation{$^d$Perimeter Institute for Theoretical Physics, 31 Caroline Street North, Waterloo, ON, N2L 2Y5, Canada}

\affiliation{$^e$Department of Chemistry and Department of Physics and Astronomy, State University of New York at Stony Brook, Stony Brook, New York 11794, USA}

\begin{abstract}
Reissner–Nordström–Anti-de Sitter (RNAdS) black holes in the extended phase space exhibit critical behavior analogous to the liquid–gas system, with critical exponents matching those of van der Waals-type phase transitions. However, the kinetics of these transitions near spinodal and critical points remain poorly understood. We demonstrate that both the autocorrelation time and the variance of trajectories increase significantly as the system approaches these special points, signaling critical slowing down. This behavior is driven by the flattening of the free energy landscape, as further confirmed by the lowest eigenvalue of the Fokker–Planck equation. Moreover, we uncover a clear dynamical crossover separating gas-like and liquid-like regimes in the supercritical region. This kinetic crossover defines the Widom line that closely matches the thermodynamic one obtained from the maxima of the isobaric heat capacity. These findings contribute to a deeper understanding of the kinetics of RNAdS black holes in the vicinity of spinodal and critical points. 
\end{abstract}

\maketitle

%\tableofcontents

\noindent 
\emph{Introduction.--} Since the discovery of Hawking radiation \cite{Hawking:1975vcx}, understanding the thermodynamic nature of black holes has become increasingly essential for constructing a quantum theory of gravity. As one of the most characteristic features of black hole thermodynamics, phase transitions of AdS black holes have been deeply illuminated by identifying the cosmological constant with thermodynamic pressure\cite{Kubiznak:2016qmn}.  RNAdS black holes in particular exhibit classical critical behavior analogous to that of the liquid-gas system, with critical exponents matching those of van der Waals-type criticality \cite{Kubiznak:2012wp}. This thermodynamic analogy is further supported by studies showing that charged AdS black holes exhibit phase structures akin to those of conventional matter \cite{Hennigar:2016xwd,Tavakoli:2022kmo}, while also offering insights into their microstructure and holographic nature \cite{Wei:2019uqg,Ahmed:2023snm}.

Recent studies have employed stochastic dynamics to investigate the kinetics of black hole phase transitions \cite{Li:2020khm,Li:2020nsy,Li:2021vdp}, by assuming that black hole state switching process takes place on the free energy landscape under both deterministic and stochastic forces \cite{Wang:2015,Fang:2019}. This approach has proven effective for analyzing first-order transitions \cite{Yang:2021ljn,Yang:2023xzv}. The liquid-gas-like nature of charged AdS black holes, which features both spinodal points for first-order phase transitions and a critical point at a second-order phase transition, makes exploring their kinetics near these points especially compelling.

For black holes in the canonical ensemble,  phase transition  kinetics are governed by a one-dimensional free energy landscape \cite{Li:2020nsy}. An essential observation is that near the spinodal or  critical points, where a single stable black hole phase splits into two branches with distinct horizon radii, the free energy landscape flattens. This may result in delayed kinetic relaxation when the black hole is perturbed from equilibrium.

Here we show that the kinetics of black hole phase transitions at both the spinodal and critical points exhibit a pronounced critical slowing down, reminiscent of  behaviour observed in ordinary matter systems \cite{Hohenberg:1977,Wissel:1984,Scheffer:2009,Scheffer:2012,Nazarimehr:2020,Yan:2023}.
We also explore these kinetics beyond the critical point. In the supercritical regime, no physical observable can distinguish between the liquid and gas phases. This feature also holds for RNAdS black holes \cite{DasBairagya:2019nyv,Zhao:2025ecg,Xu:2025jrk}, where a single stable supercritical black hole phase exists on the free energy landscape. Our simulations reveal that the supercritical regime is divided into two distinct dynamical domains: gas-like and liquid-like domains. The boundary separating these regimes, known as the Widom line, is identified as the locus of maxima in the autocorrelation time. Notably, this dynamical crossover boundary is shown to be consistent with the conventional definition of Widom line as the locus of maxima of heat capacity \cite{Xu:2005PNAS,Simeoni:2010NP,Bolmatov:2013NC,Luo:2014PRL,Gallo:2014NC,Li:2024PNAS}. Our findings thus extend the analogy between black holes and ordinary matter well beyond the critical point.

\noindent 
\emph{Generalized free energy.--}
We focus on the van der Waals-type phase transition of RNAdS black holes. By evaluating the gravitational action \cite{Li:2022oup}, the generalized free energy of the RNAdS black holes in canonical ensemble is shown to be
\begin{eqnarray}\label{Ggen}
    G(r_+)=\frac{r_+}{2}\left(1+\frac{8}{3}\pi P r_+^2 +\frac{Q^2}{r_+^2}\right)-\pi T r_+^2\;,
\end{eqnarray}
where the thermodynamic pressure $P=-\frac{\Lambda}{8\pi}$ is identified with the cosmological constant \cite{Kastor:2009wy,Dolan:2011xt}. The black hole's electric charge and the ensemble temperature are respectively denoted by $Q$ and $T$. The generalized free energy $G$ is regarded as a function of the black hole radius $r_+$, which serves as the order parameter \cite{Wei:2015iwa}; for simplicity, we henceforth use the notation $r$ to represent $r_+$.

The free energy landscape for black hole phase transitions can be formulated by considering $G$ in Eq.\eqref{Ggen}  as the thermodynamic potential. The local minima on the free energy landscape correspond to stable or metastable black hole states. The RNAdS black hole exhibits a critical point, which is determined by 
\begin{eqnarray}
    G^{(1)}(r_c)=G^{(2)}(r_c)=G^{(3)}(r_c)=0\;.
\end{eqnarray}
Here the superscript $``(n)"$ represents the $n$-th order derivative of the generalized free energy with respect to $r$. The first order derivative of the generalized free energy determines the local stationary state, while the second and third-order derivatives of the free energy are associated with the specific heat and its first-order derivative.

The critical point is \cite{Kubiznak:2012wp}
\begin{eqnarray}
    r_c=\sqrt{6}Q\;\;,\;\;T_c=\frac{\sqrt{6}}{18\pi Q}\;\;,\;\;P_c=\frac{1}{96\pi Q^2} 
\end{eqnarray}
with the coexistence line (along which first order phase transitions occur) determined by requiring that the generalized free energies of the small and large black holes are equal. Spinodal points are found by imposing that the first and second derivatives of the generalized free energy with respect to the black hole radius vanish. This indicates that the thermodynamics of the phase transition is completely governed by the free energy landscape.

\begin{figure}
  \centering
  \includegraphics[width=7cm]{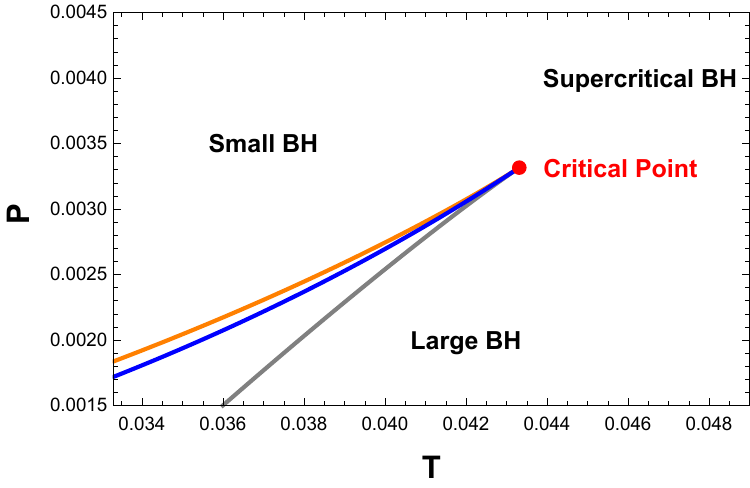}
  \caption{Phase diagram near the critical point in $T-P$ plane. The blue line represents the coexistence curve of small and large black holes, ending at the critical point (red point). The orange and gray lines are the spinodal lines.}
  \label{Phase_diagram_csd}
\end{figure}

\noindent
\emph{Kinetics of black hole phase transitions.--} The free energy landscape also offers a framework for studying the kinetics of black hole phase transitions. By assuming that the black hole state transition process occurs on the free energy landscape, the evolution of the order parameter is governed by the Langevin equation \cite{Li:2021vdp}
\begin{eqnarray}
    \frac{d^2 r}{dt^2}=-\zeta \frac{dr}{dt} -\frac{\partial G}{\partial r} + \eta(t)\;,
\end{eqnarray}
where $\zeta$ is the damping coefficient, representing the interaction between the black hole's degrees of freedom and those of the thermal bath. The stochastic force   $\eta(t)$  is assumed to be the zero-mean Gaussian white noise satisfying the fluctuation-dissipation relation
\begin{eqnarray}
    \langle \eta(t)\rangle=0\;,\;\;\;
    \langle \eta(t)\eta(s)\rangle=2\zeta T \delta(t-s)\;.
\end{eqnarray}
The free energy landscape provides the deterministic driving force, while thermal fluctuations introduce the stochastic force, together determining the dynamics of the system. 

\noindent
\emph{Critical slowing down: analytical derivation.--} Conventionally, critical slowing down is characterized by the divergence of autocorrelation time or correlation length, accompanied by enhanced fluctuations \cite{Hohenberg:1977}. The autocorrelation time $\tau$ typically follows a power-law scaling $\tau\sim |\epsilon|^{-\Delta}$, where $\epsilon=(T-T_c)/T_c$ is the reduced temperature, and $\Delta$ is the dynamic critical exponent. This slow relaxation may serve as an early warning signal of approaching spinodal line or criticality  \cite{Wissel:1984,Scheffer:2009,Scheffer:2012,Yan:2023,Nazarimehr:2020}, and underlies the formation of topological defects in continuous phase transitions, which is known as the Kibble-Zurek mechanism \cite{Kibble:1980,Zurek:1985},

First we demonstrate that the relaxation time for the black hole phase transition diverges at the critical point, which is a typical manifestation of the critical slowing down. Let us consider the deterministic time evolution of the order parameter in the overdamped limit, which is governed by the equation 
\begin{eqnarray}
    \zeta \frac{dr}{dt}=-\frac{dG}{dr}\;.
\end{eqnarray}

Far from the critical point, the free energy landscape typically takes the form of either a double-well or a single-well shape. In either case, near a stationary point, we can approximate the generalized free energy as follows 
\begin{eqnarray}\label{GFE_expansion}
    G(r)=G(r_e)+\frac{1}{2}G^{(2)}(r_e)(r-r_e)^2+\cdots \;,
\end{eqnarray}
where $r_e$ represents the radius of a stationary black hole on the landscape. This expansion uses the fact that at the stationary point, the first derivative of the generalized free energy vanishes.

For small deviations from the stationary point  (and ignoring the higher order terms in the expansion of \eqref{GFE_expansion}), the evolution of the order parameter follows an exponential form
\begin{eqnarray}\label{derm_evolution}
   r(t)- r_e=\left(r(0)-r_e \right) e^{-\gamma t}\;,
\end{eqnarray}
where $r(0)$ is the initial value of the order parameter and $\gamma=\frac{G^{(2)}(r_e)}{\zeta}$ is the decay factor. The solution indicates that the relaxation process is characterized by a timescale $\tau =1/\gamma(T)$, which is controlled by the ensemble temperature.

Near the critical point, the parabolic approximation given in Eq.\eqref{GFE_expansion} breaks down, as the derivatives of the generalized free energy with respect to the black hole radius are zero up to third order. The order parameter then evolves as
\begin{eqnarray}
    r(t)-r_c=\frac{r(0)-r_c}{\sqrt{\left(r(0)-r_c\right)^2(\beta t)+1}}\;,
\end{eqnarray}
where $\beta=\frac{G^{(4)}(r_c)}{3\zeta}=\frac{12Q^2}{r_c^5}$. Here, $\beta$ represents the strength of fourth-order fluctuations of $r$ near $r_c$, and plays a similar role to kurtosis in statistics, which quantifies the deviation from Gaussian (parabolic) behavior of the generalized free energy landscape.

Notably for  late time dynamics, the order parameter exhibits a much slower power-law decay near the critical point compared to the exponential decay observed far from the critical point. It is important to note that $\beta$ is independent of the thermodynamic pressure $P$ and the ensemble temperature $T$. This reflects the fact that the critical slowing down is a universal feature of the critical dynamics, which does not depend on external environmental parameters but is determined solely by the intrinsic properties of the critical free energy landscape.

We now consider the stochastic time evolution of the order parameter and use the autocorrelation function to predict the behaviour of the kinetics near the critical point. To this end we consider the rebound properties of the black hole state near one of the stable states. This requires that fluctuations of the order parameter remain confined to the vicinity of a single stable state. The time evolution of the order parameter in the overdamped limit with   stochastic noise can be approximated by the Langevin equation 
\begin{eqnarray}\label{L_eq_appro}
    \frac{dr}{dt}=-\frac{G^{(2)}(r_e)}{\zeta}(r-r_e)+\frac{1}{\zeta} \eta(t) \;.
\end{eqnarray}
For such a stochastic dynamics for the black hole phase transition, the critical slowing down can be revealed by the divergence of the autocorrelation time of the order parameter near the critical point.

The formal solution to Eq.\eqref{L_eq_appro} can be written as 
\begin{eqnarray}\label{Formal_solution}
    r(t)-r_e=(r(0)-r_e) e^{-t/\tau} +\frac{1}{\zeta}\int_{0}^{t} e^{-(t-t')/\tau} \eta(t')\;.
\end{eqnarray}  
Multiplying both sides of the above equation by $(r(0)-r_e)$ and taking the average, we have
\begin{eqnarray}
    \langle  (r(t)-r_e)(r(0)-r_e) \rangle = \langle (r(0)-r_e)^2 \rangle  e^{-t/\tau} \nonumber\\
    +\frac{1}{\zeta}\int_{0}^{t} e^{-(t-t')/\tau} \langle \eta(t')(r(0)-r_e) \rangle\;.
\end{eqnarray}
Taking the mean of the noise to be zero, we obtain 
\begin{eqnarray}
    \langle \left(r(t)-r_e\right)\left(r(0)-r_e\right) \rangle \sim e ^{-t/\tau}\;,
\end{eqnarray}
for the autocorrelation function of the order parameter.
which is consistent with the behavior of the deterministic time evolution given in Eq.\eqref{derm_evolution}. The autocorrelation time $\tau=\frac{\zeta}{G^{(2)}(r_e)}$ that characterizes the exponential decay is divergent at the critical point since the derivatives of generalized free energy with respect to the black hole radius vanish up to third order.

\begin{figure}
  \centering
  \includegraphics[width=4.25cm]{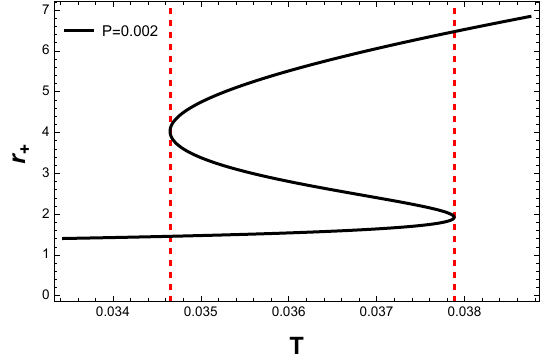}
  \includegraphics[width=4.25cm]{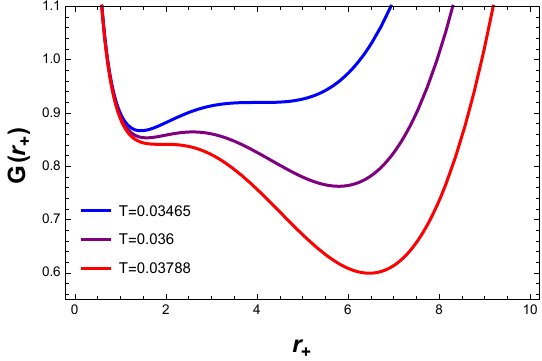}
  \caption{The spinodal diagram and the corresponding landscapes. The dashed lines represent the left and the right spinodal temperatures, at which a large black hole state emerges (blue line on the right panel) or a small black hole state disappears (red line on the right panel) on the landscape.}
  \label{spinodal_plot}
\end{figure}

\begin{figure}
  \centering
  \includegraphics[width=4.25cm]{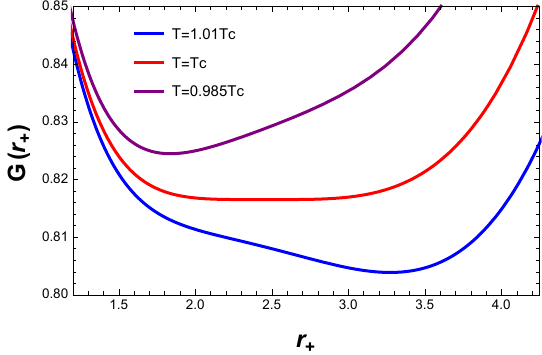}
  \includegraphics[width=4.25cm]{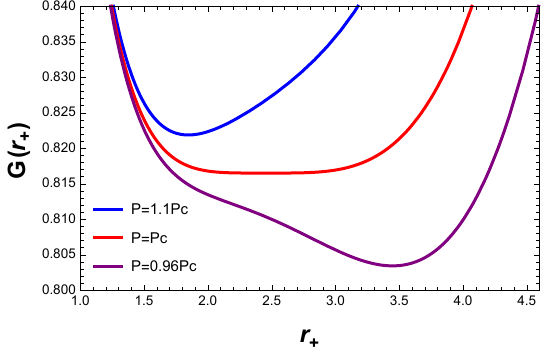}
  \caption{Landscapes with varying the ensemble temperature $T$ (left panel) or pressure $P$ (right panel). The red line represents the system  at the critical point, where a very flat potential is shown.}
  \label{Landscapes_varying_T_or_P}
\end{figure}

In fact, the generalized free energy at the spinodal point exhibits behavior analogous to that at the critical point, as can be seen in Fig.\ref{spinodal_plot} and Fig.\ref{Landscapes_varying_T_or_P}. \iffalse \jy{Are they the same point, corresponding to the same generalized free energy? }\fi Near the spinodal point, the parabolic approximation \eqref{GFE_expansion} for the generalized free energy also breaks down, as its derivatives with respect to the black hole radius vanish up to the second order. Consequently, the autocorrelation time $\tau$ also diverges, indicating a pronounced slowing-down effect in the kinetic evolution.

Summarizing, as the black hole approaches the spinodal line(s) or the critical points, its return to a stable attractor under small perturbations becomes increasingly sluggish. This makes the autocorrelation time a valuable diagnostic tool for identifying the proximity of a critical transition.

\noindent
\emph{Critical slowing down: numerical results.--} Without loss of generality, we now take $\zeta=1$ and set $Q=1$ in the following. By simulating the Langevin equation, we obtain the time series of the order parameter, from which the autocorrelation function can be statistically analyzed. The autocorrelation time $t$ is then extracted by fitting the autocorrelation function to an exponential decay form. In addition to the autocorrelation function, we also compute the variance of the trajectories and the smallest eigenvalue of the Fokker-Planck equation to further characterize the system’s dynamics \cite{Dekker_Kampen}. The variance captures the magnitude of fluctuations around the metastable states, serving as an indicator of dynamical sensitivity, while the smallest nonzero eigenvalue of the Fokker-Planck equation quantifies the relaxation timescale, thus also providing insight into the system’s approach to equilibrium. The numeric schemes are presented in the Appendix.

\begin{figure}
  \centering
  \includegraphics[width=8cm]{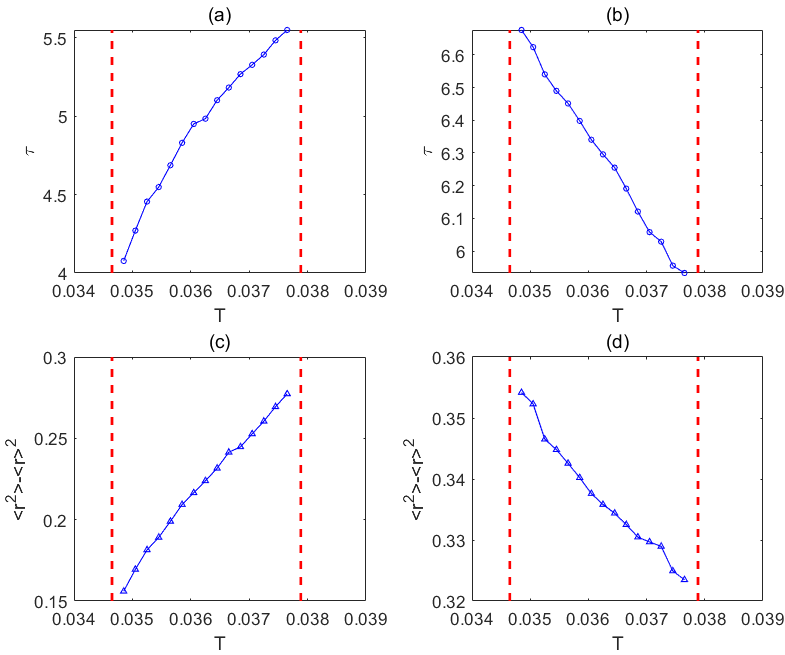}
  \caption{The correlation time $\tau$ (top figures) and the variance $\langle r^2\rangle-\langle r\rangle^2$ (bottom figures) of trajectories as the functions of $T$ for the kinetics near the spinodal points. Left/right panels show the kinetics near the small/large black hole. \iffalse\jy{Can we plot more points for Fig3 and Fig5 to make it more smooth?} \textcolor{red}{\bf Response: The current resolution is adequate, as the primary purpose of the plot is to illustrate the general trend in the correlation time and associated observables.}\fi }
  \label{bifur_tau}
\end{figure}

We begin by analyzing the kinetics near the spinodal point (see Fig.\ref{spinodal_plot}). As the temperature increases from the left spinodal temperature, the potential well associated with the small black hole gradually flattens and eventually disappears at the right spinodal temperature. Consequently, both the autocorrelation time and the variance of trajectories near the small black hole grow with increasing temperature, as shown in the left panels of Fig.\ref{bifur_tau}, clearly indicating critical slowing down.

Conversely, when the temperature decreases from the right spinodal temperature, the potential well corresponding to the large black hole progressively flattens and vanishes at the left spinodal temperature. As a result, the autocorrelation time and trajectory variance near the large black hole increase as the temperature drops, as demonstrated in the right panels of Fig.~\ref{bifur_tau}, again signalling the presence of critical slowing down.

We now analyze the kinetics near the critical point. We consider the following two distinct paths: (a) Fixing the pressure at the critical value, $P=P_c$, while varying the ensemble temperature $T$ near the critical temperature $T_c$; (b) Fixing the temperature at the critical value, $T=T_c$, while varying the pressure $P$ near the critical pressure $P_c$. The corresponding landscapes are shown in Figure \ref{Landscapes_varying_T_or_P}. The landscape maintains a single-well structure, indicating that the system supports only one stable black hole state. As the system approaches the critical point, the potential well gradually flattens, signaling the onset of the slowing-down effect.

\begin{figure}
  \centering
  \includegraphics[width=8cm]{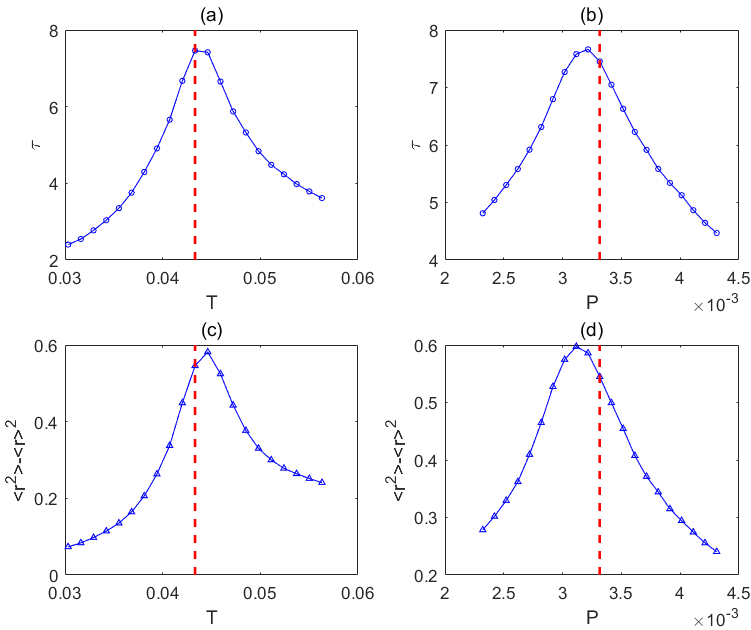}
   \caption{The auto-correlation time $\tau$ (top panels) and the variance of trajectory $\langle r^2\rangle-\langle r\rangle^2$ (bottom panels) as the functions of $T$ (left panels) or $P$ (right  panels). The vertical dashed lines represent the critical temperature or the critical pressure.}
  \label{Tau_Varying_T_or_P}
\end{figure}

\begin{figure}
  \centering
  \includegraphics[width=4.25cm]{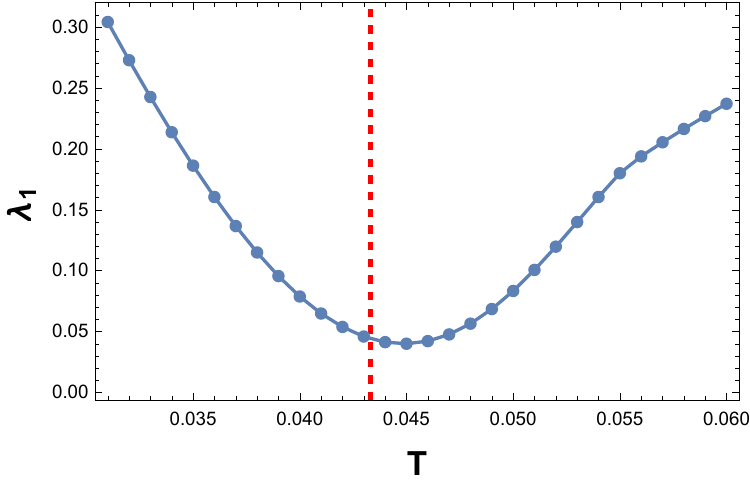}
   \includegraphics[width=4.25cm]{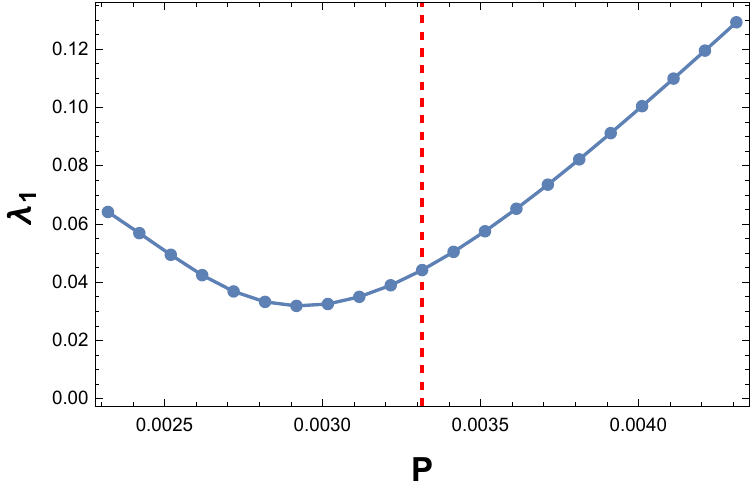}
  \caption{The smallest nonzero eigenvalue $\lambda_1$ of the Fokker-Planck equation with the fixed pressure $P=P_c$ (left panel) or the fixed ensemble temperature $T=T_c$ (right panel). The plots show the dependence of $\lambda_1$ on the temperature or the pressure. }
  \label{Eigenvalue_Varying_T_or_P}
\end{figure}

This is evidenced by the significant increase in both the autocorrelation time and the trajectory variance at the critical point, as depicted in Fig.\ref{Tau_Varying_T_or_P}. Furthermore, the smallest eigenvalue of the Fokker-Planck equation, plotted in Fig.\ref{Eigenvalue_Varying_T_or_P}, exhibits a corresponding suppression near the critical point, further reinforcing the slowing-down behavior.  \iffalse\jy{Explain why this suppression corresponds to slow down?}\textcolor{red}{\bf Response: This is because the inverse of the smallest nonzero eigenvalue determines the longest relaxation timescale of the system.}\fi Therefore, in both cases, we observe a strong signal of slowing-down effect of the kinetics near the critical point. However, the slowest kinetic behavior does not occur exactly at the critical point: in case (a), it appears slightly above the critical temperature, and in case (b), slightly below the critical pressure. \iffalse\jy{Can we explain why the peaks are not in the critical value, for $T$, peak is on the right, for $P$, peak is on the left?} \textcolor{red}{\bf I have no idea.}\fi

\noindent
\emph{Kinetic crossover in supercritical regime.--} In systems exhibiting critical behavior, the Widom line extends the concept of phase separation into the supercritical regime, beyond the critical point where no true phase transition occurs. Although the system remains in a single thermodynamic phase, the Widom line marks a transition between qualitatively different liquid-like and gas-like regimes. While the Widom line is commonly defined as the locus of maxima in thermodynamic response functions such as heat capacity, compressibility, or correlation length, its precise and universal definition remains an open question.

\begin{figure}
  \centering
  \includegraphics[width=8cm]{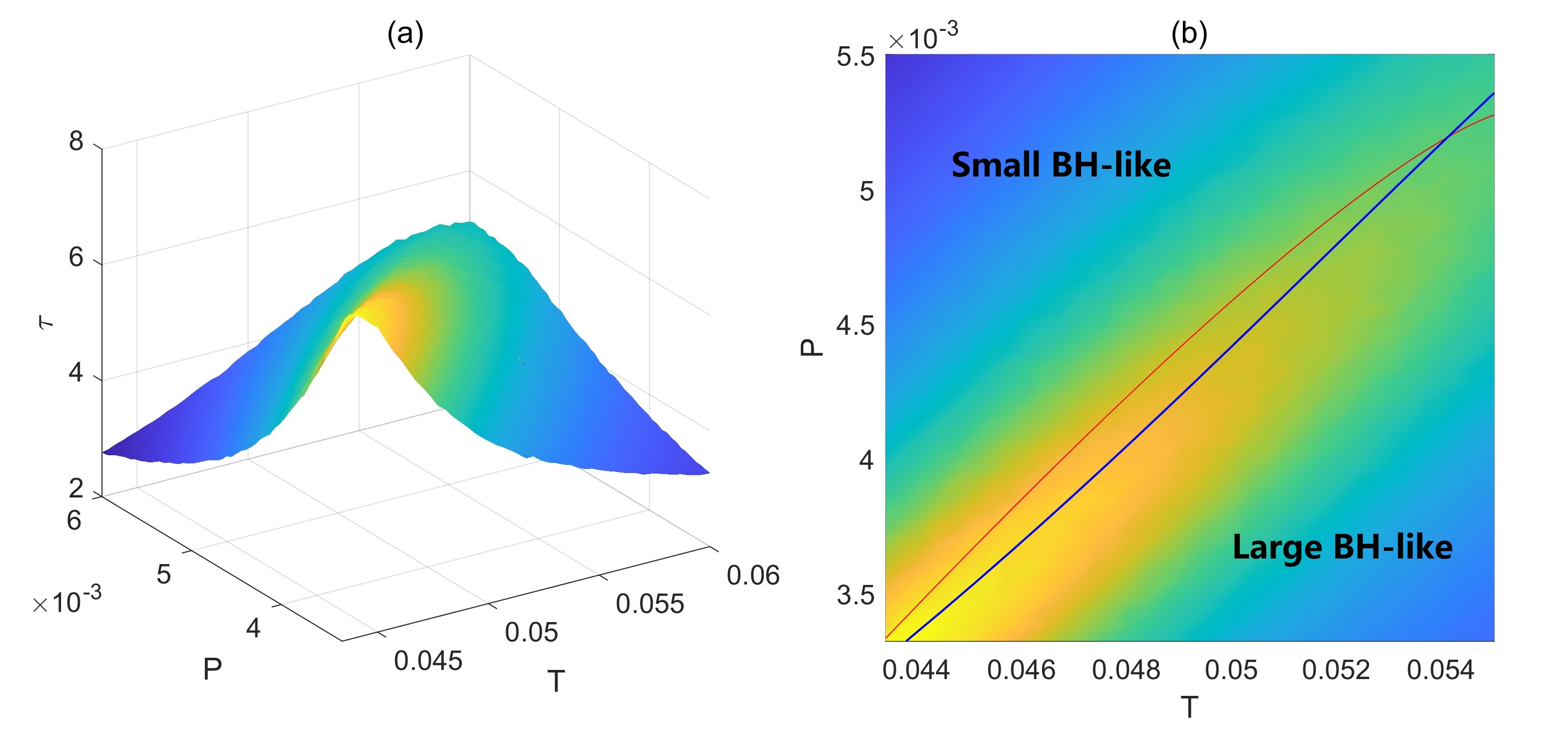}
  \caption{Left panel: 3D plot of the correlation time as the function of $T$ and $P$ in the supercritical regime; Right panel: the corresponding density plot and the Widom line. The red line denotes the locus of maxima of the isobaric heat capacity, and the blue line denotes the locus of the kinetic crossover.} 
  \label{supercritical_kinetics}
\end{figure}

\begin{figure}
  \centering
  \includegraphics[width=3.8cm]{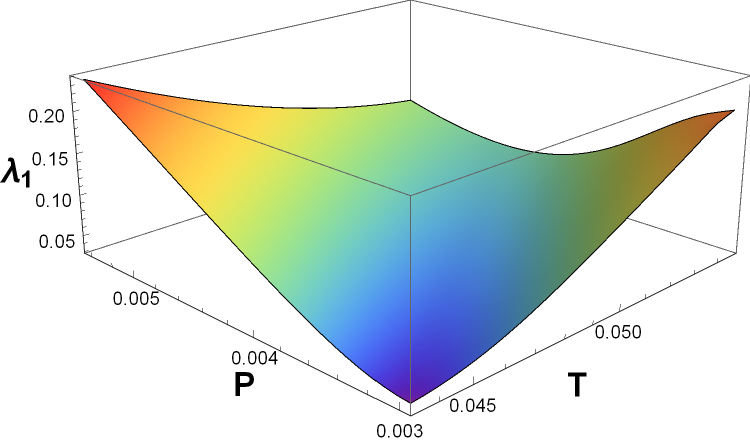}
  \includegraphics[width=3.8cm]{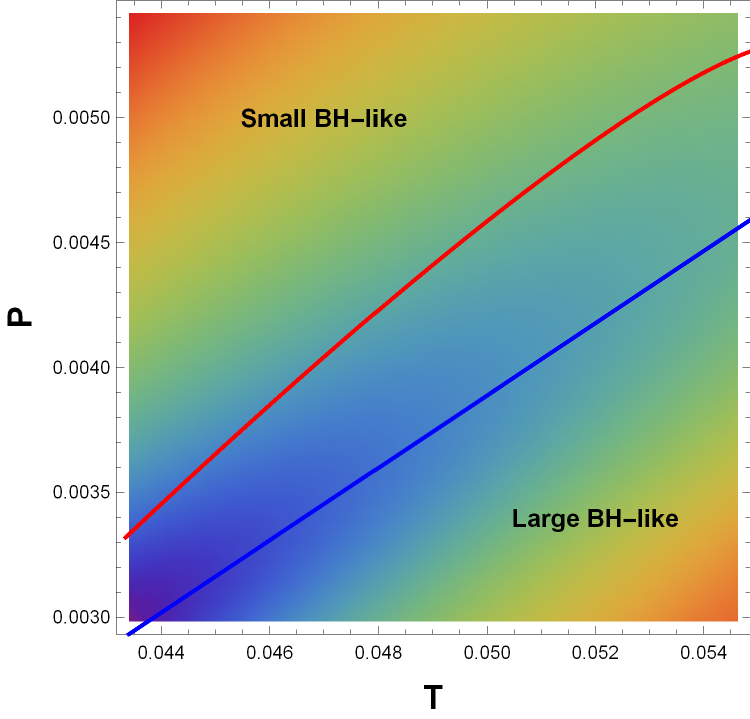}
  \caption{Left panel: 3D plot of the lowest eigenvalues for the supercritical black holes as the function of $T$ and $P$; Right panel: the corresponding density plot and the Widom line. The red line denotes the locus of maxima of the isobaric heat capacity, and the blue line denotes the locus of the crossover in the lowest eigenvalue. }
  \label{Eigenvalues_SBH}
\end{figure}

We propose that the Widom line can be identified by exploring the kinetics of black hole state transits in the supercritical regime. The left panel of Fig.~\ref{supercritical_kinetics} shows the transition kinetics characterized by the autocorrelation times, where a clear dynamical watershed emerges, separating gas-like and liquid-like behaviors. This watershed is interpreted as the Widom line. On the right panel of Fig.~\ref{supercritical_kinetics}, this kinetically determined boundary (blue line) is compared with the conventional Widom line, defined as the locus of maxima of the isobaric heat capacity (red line). We find that the two criteria are in close agreements. The lowest nonzero eigenvalues of Fokker-Planck equation for the supercritical black holes, shown in Fig.~\ref{Eigenvalues_SBH}, display the behavior analogous to autocorrelation times, reinforcing the identification of the kinetic crossover.

These observations suggest that the kinetic crossover provides a compelling criterion for distinguishing the gas-like and liquid-like phases of supercritical black holes. Unlike the conventional thermodynamic definition, which is based on the equilibrium fluctuations, the kinetic definition of the Widom line relies on the underlying dynamical processes of black hole state transitions. As such, it offers deeper insights into the physical significance of the Widom line in the black hole systems.

\noindent
\emph{Conclusion} 
We have studied the kinetic behavior of black hole phase transitions near both spinodal lines and critical points. We have demonstrated analytically the divergence of the autocorrelation time at these special points; numerically we find that both the autocorrelation time and the variance of trajectories exhibit pronounced peaks near criticality, consistent with the onset of slow dynamics. The underlying reason is from the flattening of the landscape near these special points. These findings are further supported by the suppression of the smallest eigenvalue of the Fokker–Planck equation near these special points. Notably, our results reveal that the slowing-down effect is not only present at the second-order critical point, but also appears prominently at the spinodal points of the first-order phase transition. We should emphasize that this slowing-down effect can be properly used as a predictor or early warning sign of dominance of a new black hole new state.

We also explored the kinetic behavior of supercritical black holes, where the system no longer exhibits a sharp phase boundary. Despite the absence of a thermodynamic phase transition, our results demonstrate a clear dynamical crossover between gas-like and liquid-like regimes, separated by a well-defined Widom line. This line, traditionally defined as the locus of maxima of response functions (e.g., isobaric heat capacity), is shown to closely align with the kinetic crossover extracted from dynamical observables, such as the autocorrelation time and the lowest eigenvalue. These results not only show that the Widom line provides a useful diagnostic of supercritical dynamics, but also strengthen the analogy between black hole and ordinary matter phase transitions.

Our analysis is limited to black holes in the canonical ensemble, where the free energy landscape is effectively one-dimensional. In more general settings, such as the grand canonical ensemble, the dynamics becomes richer due to the presence of multiple order parameters (e.g., black hole radius and charge \cite{Li:2023ppc, Liu:2023sbf}). The associated two-dimensional free energy landscape introduces additional degrees of freedom and complexity in the kinetics. Investigating critical slowing down and dynamical crossover in such higher-dimensional settings provides an important direction for future research and could offer deeper insights into the kinetics of black hole phase transitions.

\noindent
\textit{Acknowledgements.--} This work was supported in part by the Natural Sciences and Engineering Research Council of Canada. Research at Perimeter Institute is supported in part by the Government of Canada through the Department of Innovation, Science, and Economic Development and by the Province of Ontario through the Ministry of Colleges and Universities.

\newpage

\section*{Appendix}
\label{Appendix}

In this appendix, we outline the numerical methods employed for integrating the Langevin equation and computing the eigenvalues of the Fokker-Planck equation. \iffalse\jy{forget mention appendix in maintext?} \textcolor{red}{Response: One sentence is added in the maintext. }\fi

The Langevin equation is a stochastic differential equation of the form
\begin{eqnarray}
    dX(t) = f(X(t)) dt + g dW(t)\;,
\end{eqnarray}
where $dW(t)$ is an independent Wiener process. Let $h$ denote the time step used for numerical integration. According to the properties of the Wiener process, at each discretized time point $t_n$, we generate independent random variables $\eta(t_n)$, distributed as $\sqrt{h} \mathcal{N}(0,1)=\mathcal{N}(0,h)$, where $\mathcal{N}(a, \sigma^2)$ stands for the normal distribution with mean $a$ and variance $\sigma^2$.

One simple discretization scheme is the Heun method, which is a predictor-corrector method. Given the value of $X$ at a time $t_n$ of the discretization, we first obtain the predictor with the Euler integration scheme
\begin{eqnarray}
    \bar{x}(t_{n+1})=x(t_n)+f(x(t_n)) h + g \eta(t_n)\;,
\end{eqnarray}
where $t_{n+1}=t_n+h$. Then one can obtain $x(t_{n+1})$ as 
\begin{eqnarray}
    x(t_{n+1})&=& x(t_n)+\frac{1}{2}\left[f(x(t_n))+f(x(t_{n+1}))\right] h\nonumber\\ &&+ g \eta(t_n)\;.
\end{eqnarray}
By simulating the Langevin equation with the Heun scheme,  
we can obtain the evolutionary trajectories of the black hole state on the free energy landscape and calculate the auto-correlation function of the order parameter.  We obtain the dependence of the correlation time on the black hole parameters by adjusting them.

In addition to working with the Langevin equation, we also calculate the eigenvalues of the corresponding Fokker-Planck equation. It can be shown that the probability distribution $P(r,t)$ of the black hole state on the free energy landscape evolves according to the following partial differential equation \cite{Li:2020khm,Li:2020nsy}
\begin{eqnarray}
    \frac{\partial P(r,t)}{\partial t} =
    D \frac{\partial}{\partial r} \left\{ e^{-\beta G(r)} \frac{\partial}{\partial r} \left[e^{\beta G(r)} P(r,t) \right] \right\}\;,
\end{eqnarray}
where $D=\frac{k_BT}{\zeta}$ is the diffusion coefficient. We set  the Boltzmann constant $k_B=1$ in the numerics.

The stationary solution to the Fokker-Planck equation is given by $P_0\propto e^{-\beta G(r)}$. The eigenvalues of the Fokker-Planck equation can be calculated by casting it into the form of a self-adjoint Schrodinger equation by performing the following spectral expansion as
\begin{eqnarray}
    P(r,t)=P_0^{1/2} \psi(r) e^{-\lambda t}\;. 
\end{eqnarray}
yielding
\begin{eqnarray}
    -T \frac{d^2 \psi}{dr^2}+V(r) \psi =\lambda\psi\;,
\end{eqnarray}
for the corresponding Schrodinger equation, 
where the effective potential is given by 
\begin{eqnarray}
    V(r)=-\frac{1}{2} G^{''}(r)+\frac{1}{4T} (G')^2\;.
\end{eqnarray}
The smallest eigenvalue $\lambda_1$ characterizes the relaxation process for the black hole state, which in turn determines the timescale of the kinetics. Slow kinetics corresponds to  small values of $\lambda_1$.

We use the pseudo-spectral method to solve for the eigenvalues of the Schrodinger equation. The basic numerical strategy to solve differential equations is to transform them into algebraic equations;  linear differential equations are then reduced to  standard linear algebra problems. By discretizing the computational domain into a finite number of grid points, an analytic function can be  represented approximately by the values on the grid points. A continuous differential equation is thus transformed into a series of algebraic equations. In this process, the differential matrix is the essential object, with various discretizing strategies differing only in their choices of the differential matrix. To obtain an approximate solution to a differential equation, one typically imposes the condition that the equation holds at the grid points.

To illustrate pseudo-spectral method, let us begin with a one dimensional computation domain with $0\leq x\leq1$. An analytical function $f(x)$ can be expanded in terms of Chebyshev Polynomials $T_n(x)$ as 
\begin{eqnarray}
    f(x)=\sum_{n=0}^{N} c_n T_n(2x-1)\;,
\end{eqnarray}
where $N$ is the truncation number and $T_n(x)=\cos(n\arccos(x))$. However, we do not work with the expansion coefficients $\{c_n\}$ but instead with the values of the function on the grid points. The strategy is to discretize the domain with Chebyshev grids 
\begin{eqnarray}
    x_i=\cos\left(\frac{\pi i}{N}\right)\;,\;\;\; i=0,\cdots N\;.
\end{eqnarray}
The function $f(x)$ is then represented by the values on the grid points $\{f(x_i)\}$. The essential point is that the derivatives of the function at the grid points is given by the following linear transformation 
\begin{eqnarray}
    f'(x_i)=\sum_{j=0}^{N} (D_N)_{ij} f(x_j)\;,
\end{eqnarray}
where $D_N$ is the Chebyshev differentiation matrix with the elements given by 
\begin{eqnarray}
   (D_N)_{ij}=\left\{ \begin{array}{cc}
        \frac{2N^2+1}{6}\;, & i=j=0 \\
        \\
        -\frac{2N^2+1}{6}\;, & i=j=N \\
        \\
         -\frac{x_j}{2\left(1-x_j^2\right)}\;, & i=j\;,\;\;0<j<N \\
         \\
         \frac{\kappa_i}{\kappa_j}\frac{(-1)^{i-j}}{x_i-x_j}\;, & i\neq j \\ 
    \end{array}
    \right.
\end{eqnarray}
where the coefficient $c_i$ is defined as 
\begin{eqnarray}
    c_i=\left\{\begin{array}{cc}
    2\;,& i=0, \textrm{\ or\ } N \\
    1\;,& 0<i<N
    \end{array}
    \right.
\end{eqnarray}

Unlike the finite difference method, where the derivative at one grid is determined solely by its neighboring points, spectral differentiation instead relies on a global interpolation involving all grid points across the entire domain. Higher derivatives can be taken as the powers of first-derivative matrices, i.e. $D_N^{(n)} = (D_N)^n$. Therefore, the differential operators can be discretized by using the corresponding Chebyshev differentiation matrices.

For the generalized free energy of the RNAdS black hole, the computational domain is given by $[r_{min},+\infty]$ with $r_{min}=r_0$ being the minimal horizon radius. One method to compactify the domain is to use a coordinate transformation somewhat like the tangent function. Here, by noting the free energy is divergent for large $r$, it is reasonable to truncate the infinite computation domain to a finite one terminating at a large $r_{max}$. Typically, we  n select $r_{max}=10$. For a finite domain, the Chebyshev grids is given by 
\begin{eqnarray}
    r_i=r_{min}+\frac{\Delta r}{2}\left[\cos\left(\frac{\pi i}{N}\right)+1\right]\;, \;\;\;0\leq i\leq N\;,
\end{eqnarray}
with $\Delta r=r_{max}-r_{min}$. The Chebyshev differentiation matrix should be multiplied by a factor of $\frac{2}{\Delta r}$.    

The Schrodinger equation can be written in the form 
\begin{eqnarray}
    \mathcal{L}\psi=\lambda\psi\;,
\end{eqnarray}
with the operator $\mathcal{L}=-T\frac{d^2}{dr^2}+V(r)$. By using the differentiation matrices, the operator $\mathcal{L}$ can be discretized as a matrix
\begin{eqnarray}
    \mathcal{L}_{ij}=-T (D_N^2)_{ij}+V(r_i) \delta_{ij}\;.
\end{eqnarray}

\begin{figure}
  \centering
  \includegraphics[width=4.25cm]{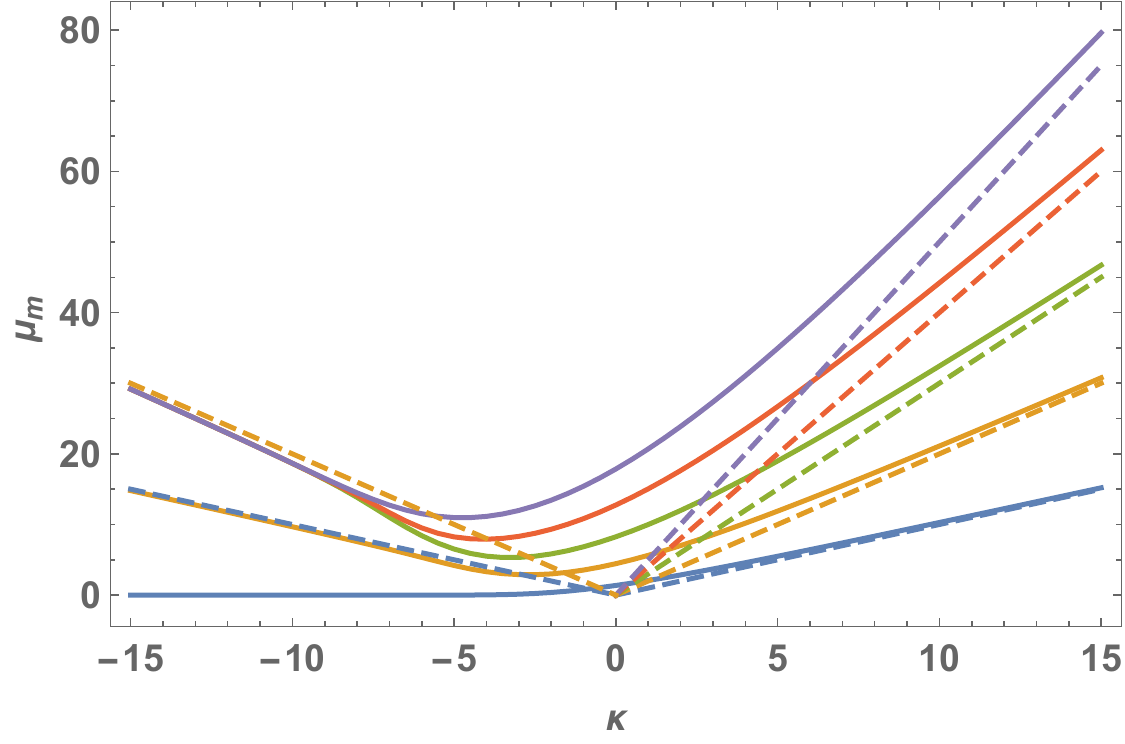}
  \includegraphics[width=4.25cm]{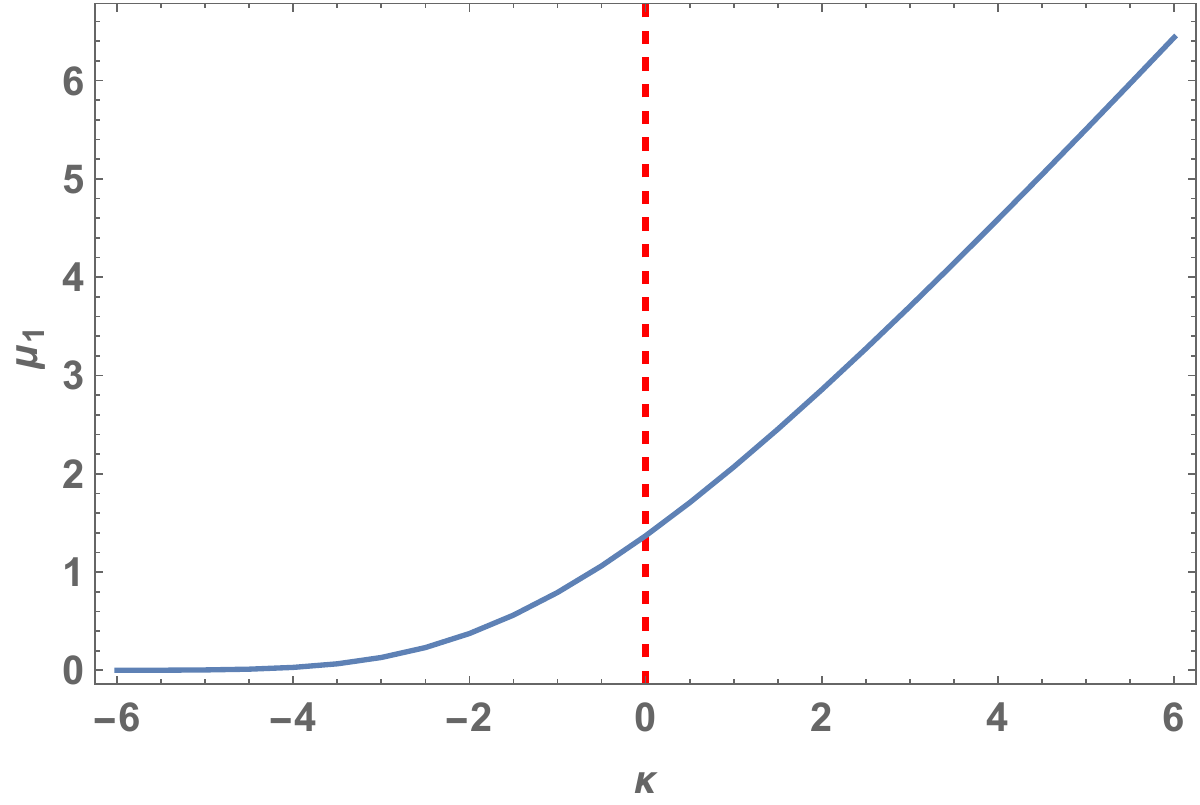}
  \caption{Eigenvalues for the Fokker-Planck equation for the symmetric double well potential \eqref{qrtic} considered in \cite{Dekker_Kampen}. Here, $\kappa$ is the pump parameter and $\mu_m$ denotes the $m$-th non-zero eigenvalue. In the left panel, the dashed lines represent the asymptotic eigenvalues. In the right panel, the vertical dashed line represents the critical value of $\kappa$.} 
  \label{Eigenvalue_double_well}
\end{figure}

In this way, the eigenvalues of Fokker-Planck equation can be recast into the eigenvalues of the linearized problem, which can be easily obtained by using the \textit{Eigenvalues} command in \textbf{Mathematica} to solve this equation numerically.

To check the robustness of this numerical scheme, we reproduce the eigenvalues of the Fokker-Planck equation considered by Dekker and Kampen in \cite{Dekker_Kampen}. These authors considered a diffusion process in a quartic potential given by
\begin{eqnarray}\label{qrtic}
    U(x)=\frac{1}{2}\kappa x^2+\frac{1}{4}x^4\;,
\end{eqnarray}
where $\kappa$ is a pump parameter. For $\kappa > 0$, the potential has a single minimum at $x = 0$. When $\kappa < 0$, the potential develops a double-well structure with two minima located at $x = \pm\sqrt{-\kappa}$ and a local maximum at $x = 0$. The critical point corresponds to $\kappa=0$. The numerical results are presented in Fig.\ref{Eigenvalue_double_well}. When $\kappa\rightarrow \infty$, the potential asymptotically approaches that of a harmonic oscillator, and the eigenvalues tend to $\mu_m=m\kappa (m=0,1,2,\cdots)$. In the opposite limit $\kappa\rightarrow -\infty$, the asymptotic eigenvalues are given by 
\begin{eqnarray}
    \mu_0=\mu_1=0\;,\;\;\;\mu_2=|\kappa|\;,\;\;\;\nonumber\\
    \mu_3=\mu_4=\mu_5=2|\kappa|\;,\cdots
\end{eqnarray}
The numerical results are found to be in good agreement with these asymptotic values.

\end{document}